\documentclass{article}
\usepackage[T1]{fontenc} 
\usepackage[utf8]{inputenc} 
\usepackage[bookmarks=false]{hyperref}
\usepackage{ismir,amsmath,cite,url}
\usepackage{graphicx}
\usepackage{color}
\usepackage{subcaption}
\usepackage{booktabs}
\usepackage{cleveref}
\usepackage{amssymb}
\usepackage{amsmath}
\usepackage{stmaryrd}

\title{Impact of time and note duration tokenizations on deep learning symbolic music modeling}

\multauthor
{Nathan Fradet$^{1,2}$ \hspace{1cm} Nicolas Gutowski$^3$ \hspace{1cm} Fabien Chhel$^{3,4}$ \hspace{1cm} Jean-Pierre Briot$^1$} {
    $^1$ Sorbonne University, CNRS, LIP6, F-75005 Paris\\
    $^2$ Aubay, Boulogne-Billancourt, France\\
    $^3$ University of Angers, LERIA, 49000 Angers, France\\
    $^4$ ESEO-TECH / ERIS, 49100 Angers, France\\
{\tt\small nathan.fradet@lip6.fr}
}

\def\authorname{F. Author, S. Author, and T. Author}

\begin{document}

\maketitle
\begin{abstract}
  Symbolic music is widely used in various deep learning tasks, including generation, transcription, synthesis,
  and Music Information Retrieval (MIR). It is mostly employed with discrete models like Transformers, which require music to be tokenized, i.e., formatted into sequences of distinct elements called tokens. Tokenization can be performed in different ways.
  As Transformers can struggle at reasoning, but capture more easily explicit information, it is important to study how the way the information is represented for such models impact their performances.
  In this work, we analyze the common tokenization methods and experiment with time and note duration representations. We compare the performances of these two impactful criteria on several tasks, including composer and emotion classification, music generation, and sequence representation learning. We demonstrate that explicit information leads to better results depending on the task.
\end{abstract}

\section{Introduction}\label{sec:tok_introduction}

Most tasks involving using deep learning with symbolic music \cite{briot_deep_2020} are performed with discrete models, such as Transformers \cite{attention_is_all_you_need}.
To use these models, the music must first be formatted into sequences of distinct elements, commonly called tokens. For instance, a token can represent a note attribute or a time event. 
The set of all known tokens is commonly called the vocabulary, and each token is associated to a unique integer id. These ids are used as input and output of models.


Compared to text, tokenizing music provides greater flexibility, as a musical piece can be played by different instruments and composed of multiple simultaneous notes, each having several attributes to represent. As a result, it is necessary to serialize these elements along the time dimension. To achieve this, researchers have developed various methods of tokenizing music \cite{oore_midilike_2018, huang_remi_2020, zeng2021musicbert, miditok2021}. 

While these works present model performance comparisons between tokenizations, their main differences or similarities are not always clearly stated. Moreover, they mostly focus on music generation, for which evaluations are performed on results obtained autoregressively, which accumulates biases \cite{nucleus_sampling} and is arguably difficult to evaluate \cite{on_the_eval_music}, rather than music modeling more broadly.
Yet, Transformer models struggle at reasoning, i.e. making logical deduction based on information points in the input data \cite{helwe2021reasoning, zhang-etal-2023-survey-efficient}, but perform tasks better when fed with explicit information and instructions \cite{zhou2023large}. In the case of symbolic music, it is thus important to study how the ways the music information is represented impact model performances.

This paper's primary contribution is a thorough and well-designed comparison of common tokenization techniques. Our focus is on two critical aspects: the representation of time and note duration. We believe that they are significant and impactful design choices for any music tokenization approach. Through experiments on composer classification, emotion classification, music generation, and sequence representation, we demonstrate that these design choices produce varying results depending on the task, model type, and inference process. Autoregressive generation benefits from explicit note duration and time shift tokens, while explicit note offset is more discriminating better suited for contrastive learning approaches.

We present next the related works, followed by an analysis of music tokenization, experimental results, and finally a conclusion. The source code is publicly available. \footnote{\href{https://github.com/Natooz/music-modeling-time-duration}{https://github.com/Natooz/music-modeling-time-duration}}



\section{Decomposing music tokenization}\label{sec:tok_decomposing_tokenization}

\subsection[short]{Related works}

Early works using discrete models for symbolic music, such as DeepBach \cite{deepbach2017hadjeres} or FolkRNN \cite{folkrnn2015sturm}, rely on specific tokenizations often tied to their training data. Since then, researchers introduced more general representations applicable to any kind of music. The most commonly used are \textit{Midi-Like} \cite{oore_midilike_2018} and \textit{REMI} \cite{huang_remi_2020}. The former tokenizes music by representing tokens as the same types of events from the MIDI protocol, while the latter represents time with \textit{Bar} and \textit{Position} tokens and note durations with explicit \textit{Duration} tokens. Additionally, \textit{REMI} includes tokens with additional information such as chords and tempo.

More recently, researchers have focused on improving the efficiency of models with new tokenizations techniques: \textit{Compound Word} \cite{cpword2021}, \textit{Octuple} \cite{zeng2021musicbert} and \textit{PopMAG} \cite{popmag2020} merge embedding vectors before passing them to the model; 2) LakhNES \cite{lakhnes2019} and \cite{MuseNet}, SymphonyNet \cite{symphonynet} and \cite{bpe-for-symbolique-music2023} use tokens combining several values, such as pitch and vocabulary.


\subsection{Music tokenization design}

When analyzing the possible designs of music tokenization, we can distinguish seven key dimensions:
\begin{itemize}
	\item \textbf{Time}: Type of token representing time, either \textit{TimeShift} indicating time movements, or \textit{Bar} and \textit{Position} indicating new bars and the positions of the notes within them. We can also consider the unit of \textit{Time-Shift} tokens, either in beats or in seconds.\footnote{In this paper we only treat of the beat unit. The MIDI protocol represents time in \textit{tick} unit, which value is proportional to the time division (in ticks per beat) and tempo. Hence, working with seconds would require a conversion from ticks.}
	\item \textbf{Notes duration}: How notes durations are represented, with either \textit{Duration} or \textit{NoteOff} tokens.
	\item \textbf{Pitch}: Most works use tokens representing absolute pitch values, although recent work shed light on the expressiveness gain of representing as intervals instead \cite{kermarec_pitch_interval};
	\item \textbf{Multitrack representation}: The representation of several music tracks in a sequence, i.e., how are the notes linked to their associated track.
	\item \textbf{Additional information}: Any additional information such as chords, tempo, rests, note density. Velocity can also falls in this category;
	\item \textbf{Downsampling}: How "continuous-like" features are downsampled into discrete sets, e.g. the 128 velocity values reduced to 16 values;
  \item \textbf{Sequence compression}: Methods to reduce the sequence lengths, such as merging tokens and embedding vectors.
\end{itemize}

\begin{table}
  \resizebox{\columnwidth}{!}{
  \begin{tabular}{lcccc}
    \toprule
     & \multicolumn{2}{c}{Time} & \multicolumn{2}{c}{Note duration} \\
    \cmidrule(l){2-3} \cmidrule(l){4-5}
    Tokenization & \texttt{TimeShift} & \texttt{Bar} + \texttt{Pos.} & \texttt{Duration} & \texttt{NoteOff} \\
    \midrule
    \textit{MIDI-Like} \cite{oore_midilike_2018} & $\surd$ & - & - & $\surd$ \\
    \textit{REMI} \cite{huang_remi_2020} & - & $\surd$ & $\surd$ & - \\
    \textit{Structured} \cite{pia2021hadjeres} & $\surd$ & - & $\surd$ & - \\
    \textit{TSD} \cite{bpe-for-symbolique-music2023} & $\surd$ & - & $\surd$ & - \\
    \textit{Octuple} \cite{zeng2021musicbert} & - & $\surd$ & $\surd$ & - \\
    \bottomrule
  \end{tabular}}
  \caption{Time and note duration representations of common tokenizations. \texttt{Pos.} stands for Position.}
  \label{tab:tok_decomposing_tokenization}
\end{table}

As time and note duration can both be represented in two different ways, existing tokenizations can be easily classified based on these dimensions, as shown in \cref{tab:tok_decomposing_tokenization}. However, other dimensions offer a broader spectrum of potential designs.

For instance multitrack can be represented by \texttt{Program} tokens\footnote{Following the conventional programs from the MIDI protocol.} preceding notes as in FIGARO \cite{vonrutte2022figaro}, distinct tracks sequences separated by \texttt{Program} tokens as in MMM \cite{mmm2020}, combined note and instrument tokens as LakhNes \cite{lakhnes2019} and MuseNet \cite{MuseNet}, or merging \texttt{Program} embeddings with the associated note tokens (MMT \cite{dong2023mmt}, MusicBert \cite{zeng2021musicbert}). One could even infer each sequence separately and lately model their relationships with operations aggregating their hidden states an in ColBERT \cite{colbert2020}.

The MIDI protocol supports a set of effects and metadata that can also be represented when tokenizing symbolic music, such as tempo, time signature, sustain pedal or control changes. Some works also include explicit \texttt{Chord} tokens, detected with rule-based methods. Nevertheless, only a few works experimented with such additional tokens so far (\cite{huang_remi_2020, sustain2021ismir}).


Previous works have mainly compared tokenization strategies by evaluating models with automatic and sometimes subjective (human) metrics, but often do not proceed to comparisons between the ways to represent one of the dimensions we introduced previously. \cite{huang_remi_2020} compared results for the generation task, for the use of \texttt{Bar} and \texttt{Position} tokens versus \texttt{TimeShift} in seconds and beats.


To the best of our knowledge, no comprehensive work and empirical analysis have fairly compared these possible tokenization choices. Conducting such an assessment would require an extensive survey. In this paper, we specifically focus on the time and note duration representations, as they are the two main characteristics present in every tokenization.

We want to highlight the importance of the explicit information carried by the token types, as they directly impact the performances of models.
\texttt{TimeShift} tokens represent explicit time movements, and especially the time distances between successive notes. On the other hand, \texttt{Bar} and \texttt{Position} tokens bring explicit information on the absolute positions (within bars) of the notes, but not the onset distances between notes. One could assume that the former might help to model melodies, and the latter rhythm and structure.
For note duration, \texttt{Duration} tokens intuitively express the absolute durations of the notes, while \texttt{NoteOff} tokens explicitly indicates the offset times. With \texttt{NoteOff}, a model would have to model note durations from the combinations of previous time tokens.

Our experiments aim to demonstrate the impact of different combinations of time and note duration tokens on model performance and which combinations are suitable for different tasks. Next, we introduce our methodology.

\section{Methodology}\label{sec:tok_methodology}

\subsection{Models and trainings}\label{subsec:tok_methodology_model}

For all experiments, we use the Transformer architecture \cite{attention_is_all_you_need}, with the same model dimensions: 12 layers, with dimension of 768 units, 12 attention heads and inner feed-forward layers of 3072.

For classification and sequence representation, it is first pretrained on 100k steps and a learning rate of $10^{-4}$, then finetuned on 50k steps and a learning rate of $3\times10^{-5}$, with a batch size of 48 examples. An exception is made for the EMOPIA dataset, for which we set 30k pretraining steps and 15k finetuning steps, as it is fairly small. These models are based on the BERT \cite{bert} implementation of the Transformers library \cite{wolf-etal-2020-transformers}. We use the same pretraining than the original BERT: 1) from 15\% of the input tokens, 80\% is masked with a special \texttt{MASK} token, and 20\% is randomized; 2) half of the inputs have 50\% of their tokens (starting from the end) shuffled and separated with a special \texttt{SEP} token, and the model is trained to detect if the second part is the next of the first.

For generation, the model is based on the GPT2 implementation of the Transformers library \cite{wolf-etal-2020-transformers}: it uses a causal attention mask, so that for each element in the sequence, the model can only attend to the current and previous elements. The training is performed with teacher forcing, the cross-entropy loss is defined as: $\ell = - \sum_{t=1}^n \log p_\theta(x_t \lvert \mathbf{x}_{\leq n})$.

All trainings are performed on V100 GPUs, using automatic mixed precision \cite{mixed_precision_training}, the Adam optimizer \cite{Adam} with $\beta_1 = 0.9$, $\beta_2=0.999$ and $\epsilon = 10^{-8}$, and dropout, weight decay and a gradient clip norm of respectively $10^{-1}$, $10^{-2}$ and $3$. Learning rates follow a warm-up schedule: they are initially set to 0, and increase to their default value during the first 30\% of training, then slowly decrease back to 0. 

10\% of the data is used for validation during training, and 15\% to test models. Inputs contains 384 to 512 tokens, and begin with a \texttt{BOS} (Beginning of Sequence) token and end with a \texttt{EOS} (End of Sequence) one.

\subsection{Tokenizations}\label{subsec:tok_methodology_data}

We investigate here the four combinations of possible time and note duration representation. In the results, we refer to them as \textit{TS} (\texttt{TimeShift}), \textit{Pos} (\texttt{Position}), \textit{Dur} (\texttt{Duration}) and \textit{NOff} (\texttt{NoteOff}). It is worth noting that \textit{TS} + \textit{Dur} is equivalent to \textit{TSD} \cite{bpe-for-symbolique-music2023} and \textit{Structured} \cite{pia2021hadjeres}, \textit{TS} + \textit{NOff} is equivalent to \textit{MIDI-Like} \cite{oore_midilike_2018}, and \textit{Pos} + \textit{Dur} is equivalent to \textit{REMI} (without additional tokens for chords and tempo).

We apply different resolutions for \texttt{Duration} and \texttt{TimeShift} token values: those up to one beat are downsampled to 8 samples per beat (spb), those from one to two beats to 4 spb, those from two to four beats to 2 spb, and those from four to eight beats to 1 spb. Thus, short notes are represented more precisely than longer ones.
\texttt{Position} tokens are downsampled to 8 spb, resulting in 32 different tokens as we only consider the 4/* time signature. This allows to represent the $16^{th}$ note.
We only consider pitches within the recommended range for piano (program 0) specified in the General MIDI 2 specifications\footnote{Available on the \href{https://www.midi.org/specifications-old/}{MIDI Manufacturers Association website}.}: 21 to 108. We then deduplicate all duplicated notes. Velocities are downsampled to 8 distinct values. No additional token (e.g., \texttt{Chord}, \texttt{Tempo}) is used.

We perform data augmentation by creating variations of the original data with pitches increased and decreased by two octaves, and velocity by one value. Finally, following \cite{bpe-for-symbolique-music2023}, we use Byte Pair Encoding to build the vocabularies up to 2k tokens for generation and 5k for other tasks. All these preprocessing and tokenization steps were performed with MidiTok \cite{miditok2021}.

\section{Generation}\label{sec:tok_generation}

\begin{figure*}
  \centering

  \begin{subfigure}{.195\textwidth}
    \centering
    \includegraphics[width=\textwidth]{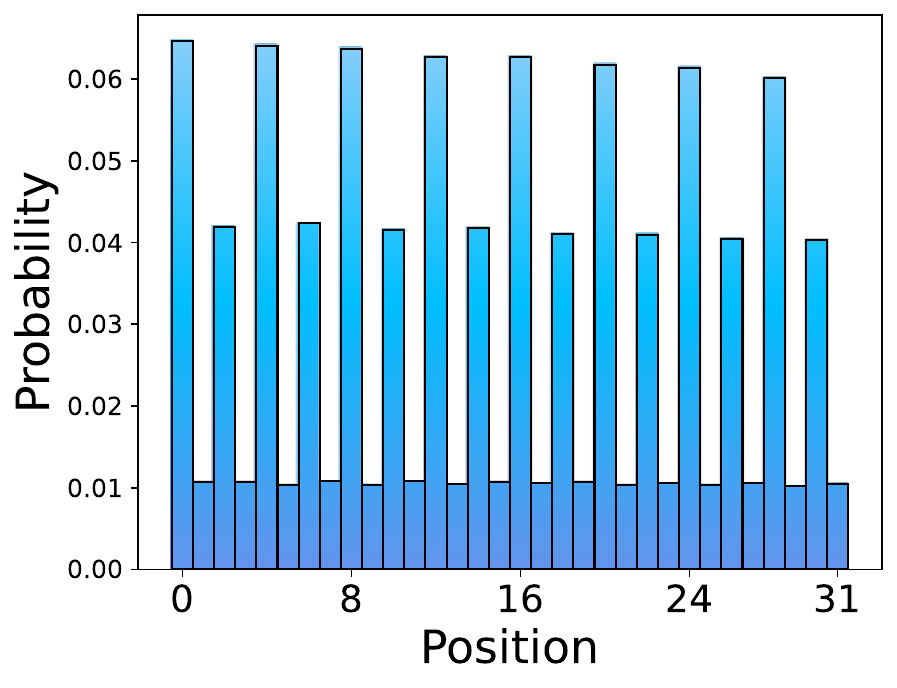}
  \end{subfigure}
  \begin{subfigure}{.195\textwidth}
    \centering
    \includegraphics[width=\textwidth]{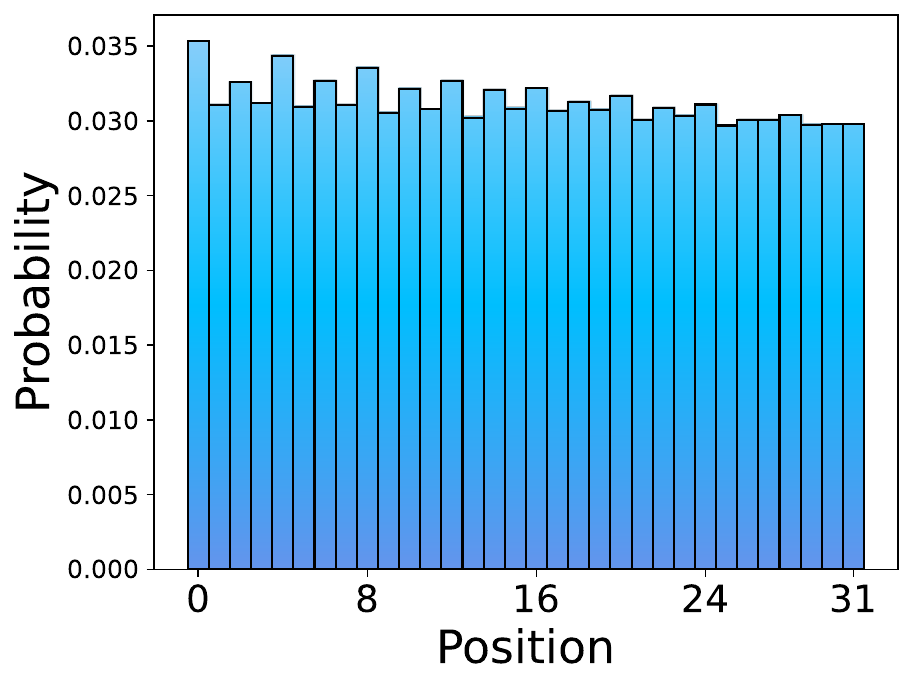}
  \end{subfigure}
  \begin{subfigure}{.195\textwidth}
    \centering
    \includegraphics[width=\textwidth]{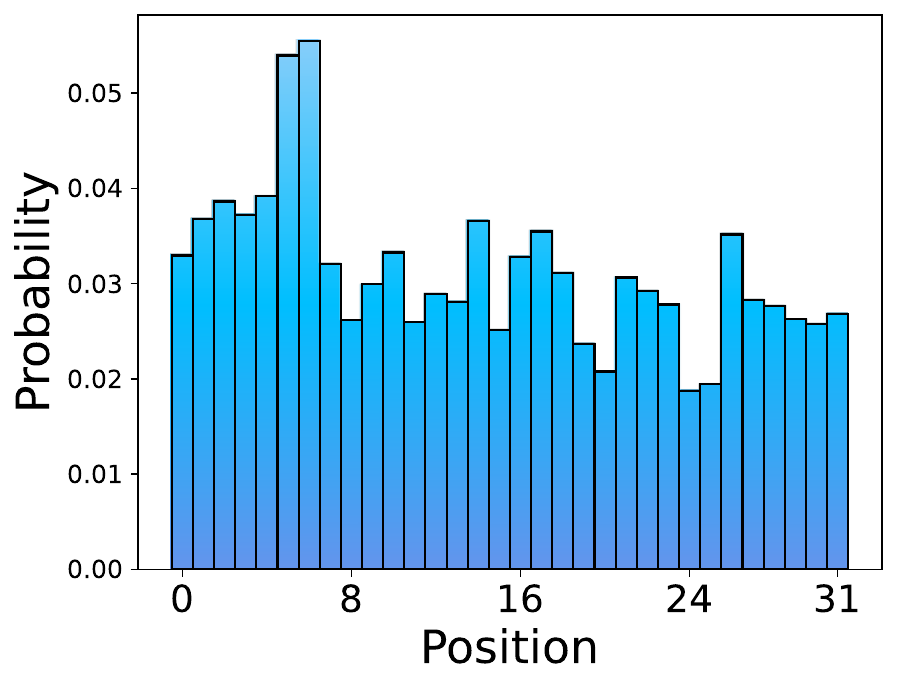}
  \end{subfigure}
  \begin{subfigure}{.195\textwidth}
    \centering
    \includegraphics[width=\textwidth]{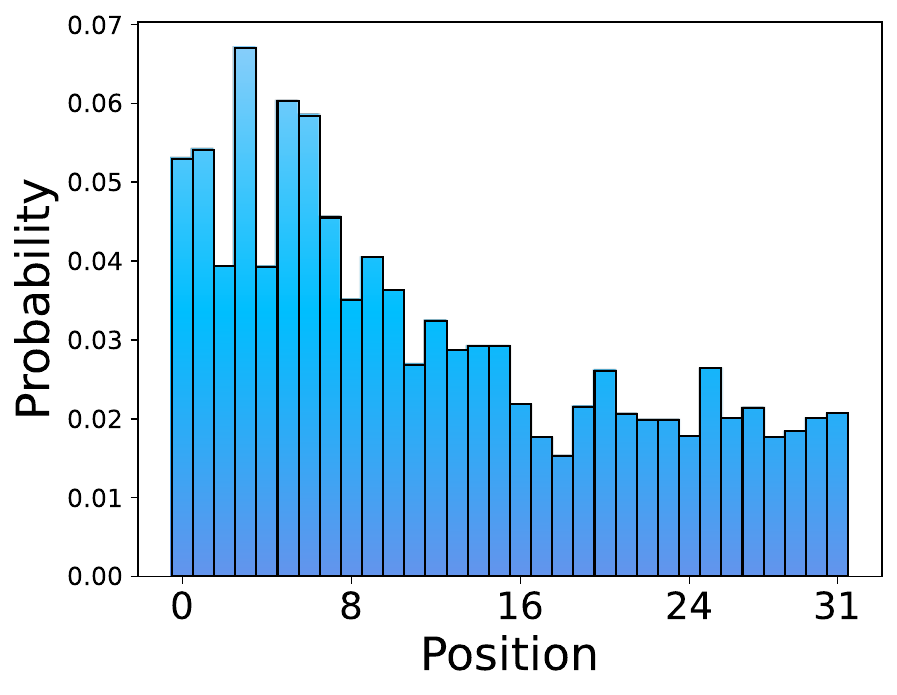}
  \end{subfigure}
  \begin{subfigure}{.195\textwidth}
    \centering
    \includegraphics[width=\textwidth]{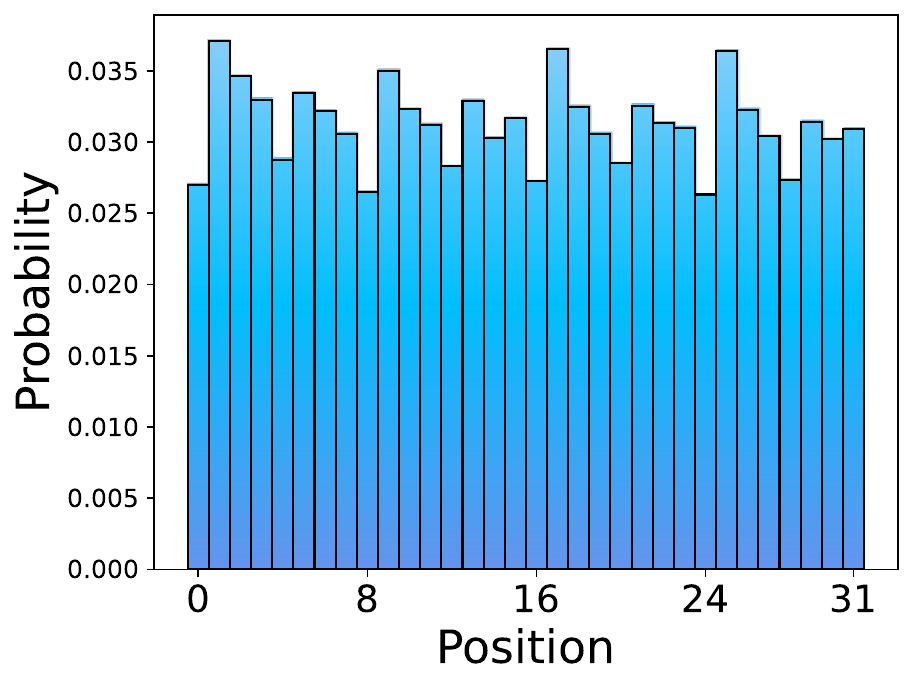}
  \end{subfigure}

  \begin{subfigure}{.195\textwidth}
    \centering
    \includegraphics[width=\textwidth]{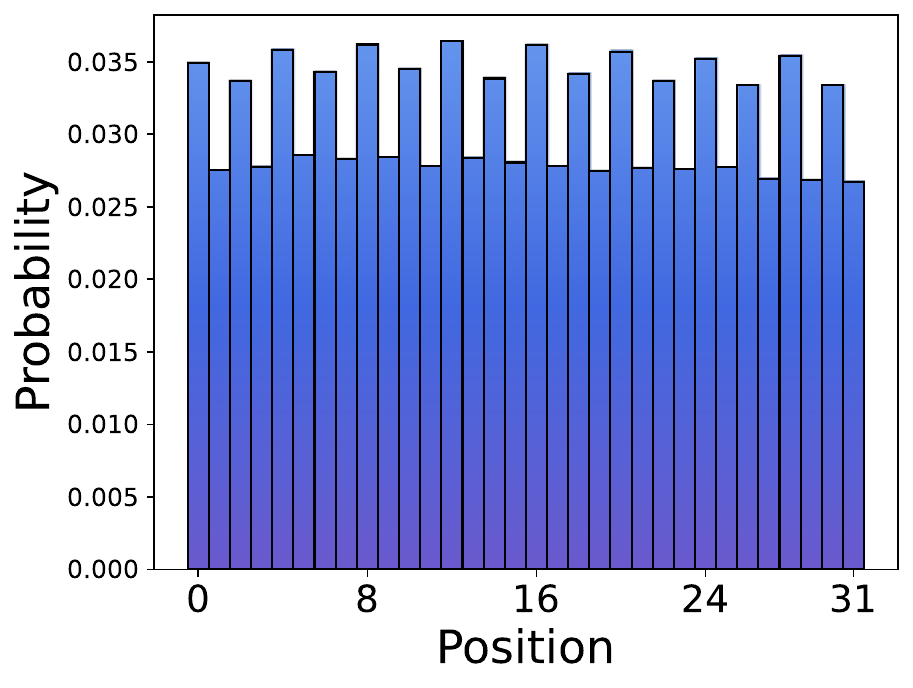}
  \end{subfigure}
  \begin{subfigure}{.195\textwidth}
    \centering
    \includegraphics[width=\textwidth]{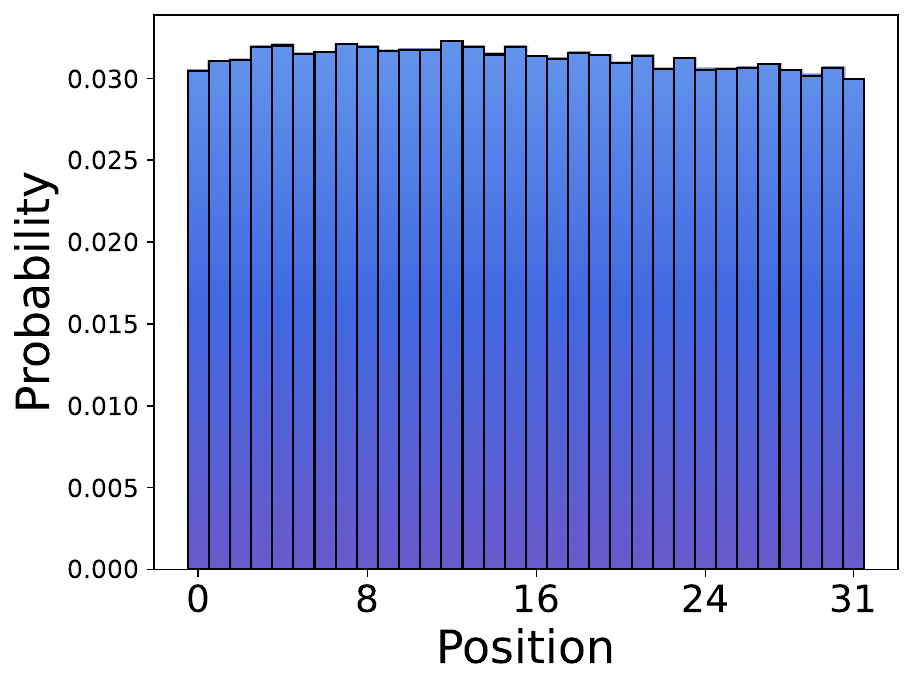}
  \end{subfigure}
  \begin{subfigure}{.195\textwidth}
    \centering
    \includegraphics[width=\textwidth]{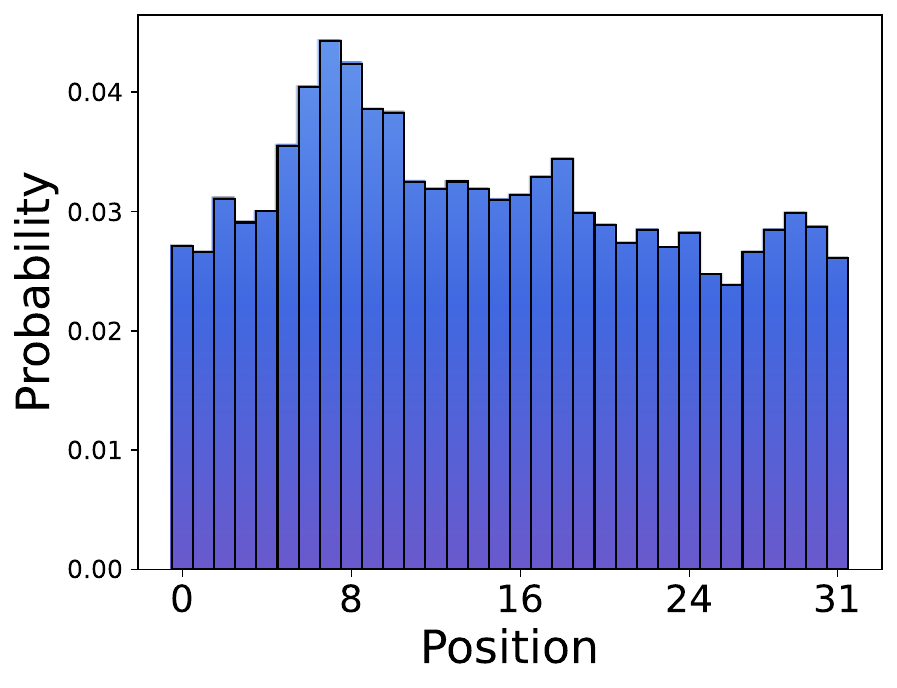}
  \end{subfigure}
  \begin{subfigure}{.195\textwidth}
    \centering
    \includegraphics[width=\textwidth]{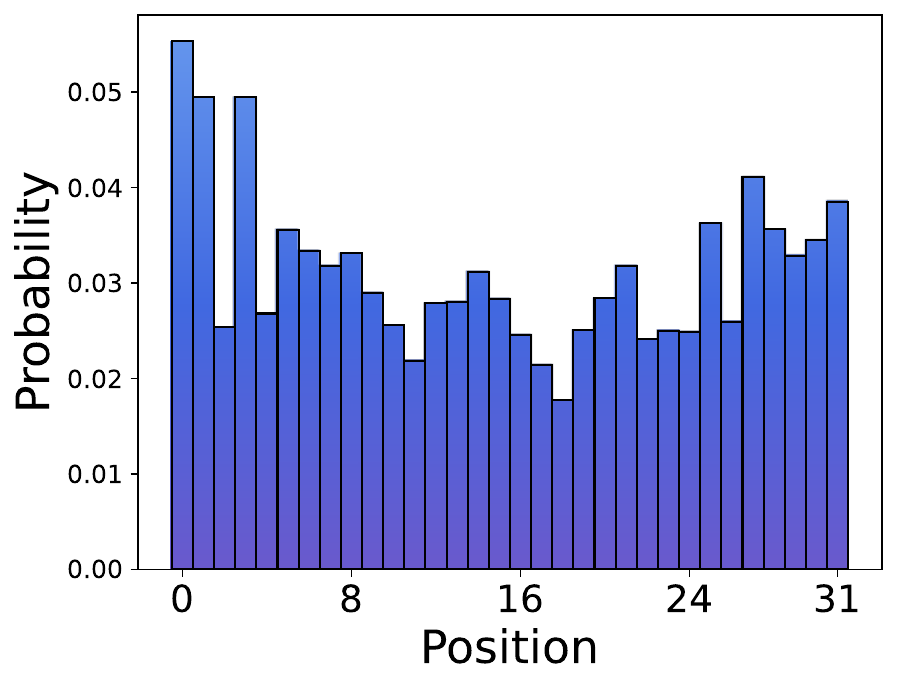}
   \end{subfigure}
   \begin{subfigure}{.195\textwidth}
    \centering
    \includegraphics[width=\textwidth]{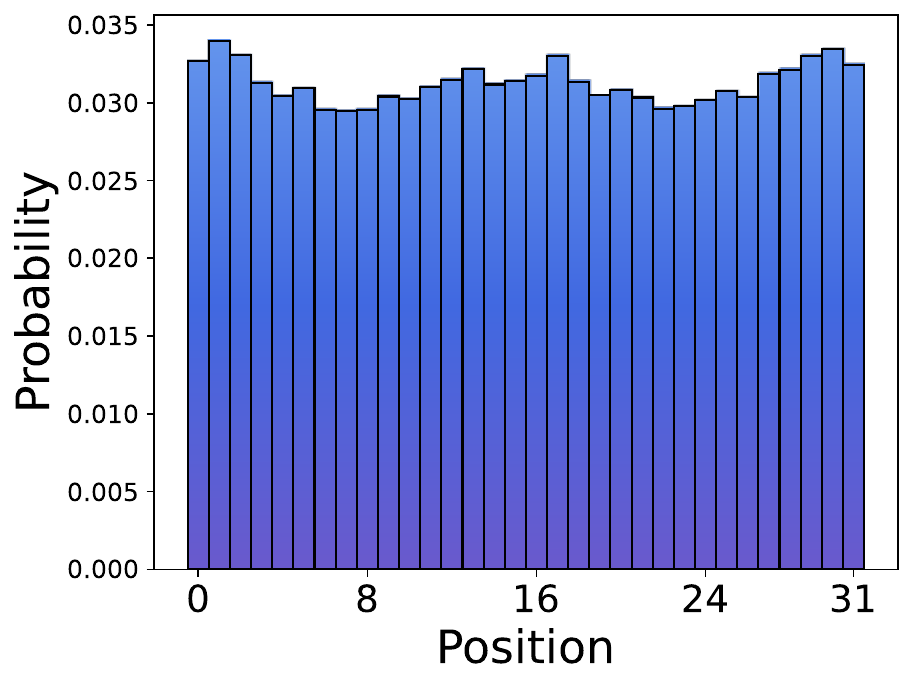}
  \end{subfigure}

  \begin{subfigure}{.195\textwidth}
    \centering
    \includegraphics[width=\textwidth]{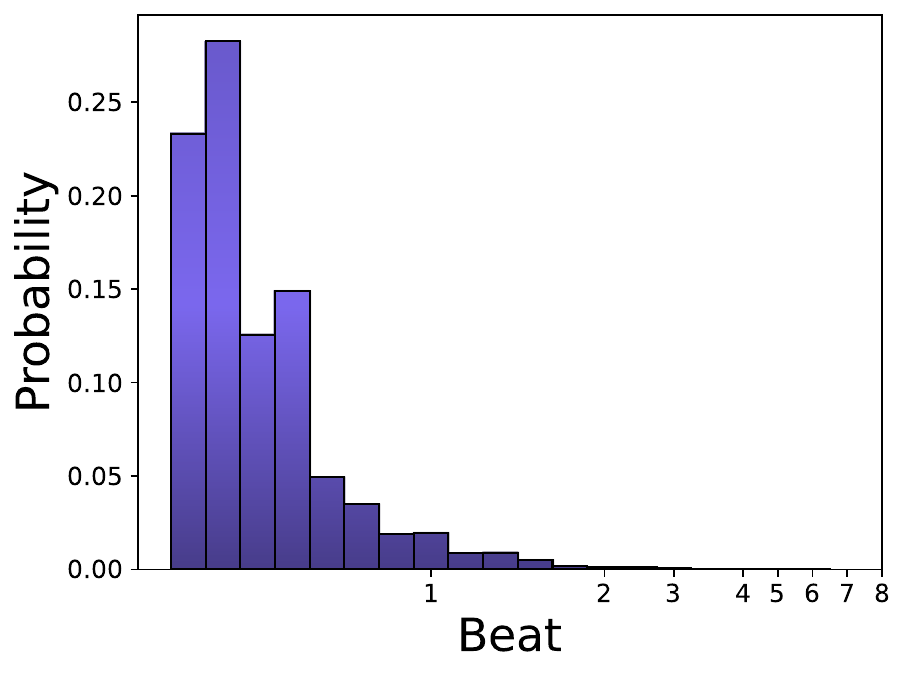}
    \vspace{-1.5\baselineskip}
    \caption[]{\textit{Ts} + \textit{Dur}}
  \end{subfigure}
  \begin{subfigure}{.195\textwidth}
    \centering
    \includegraphics[width=\textwidth]{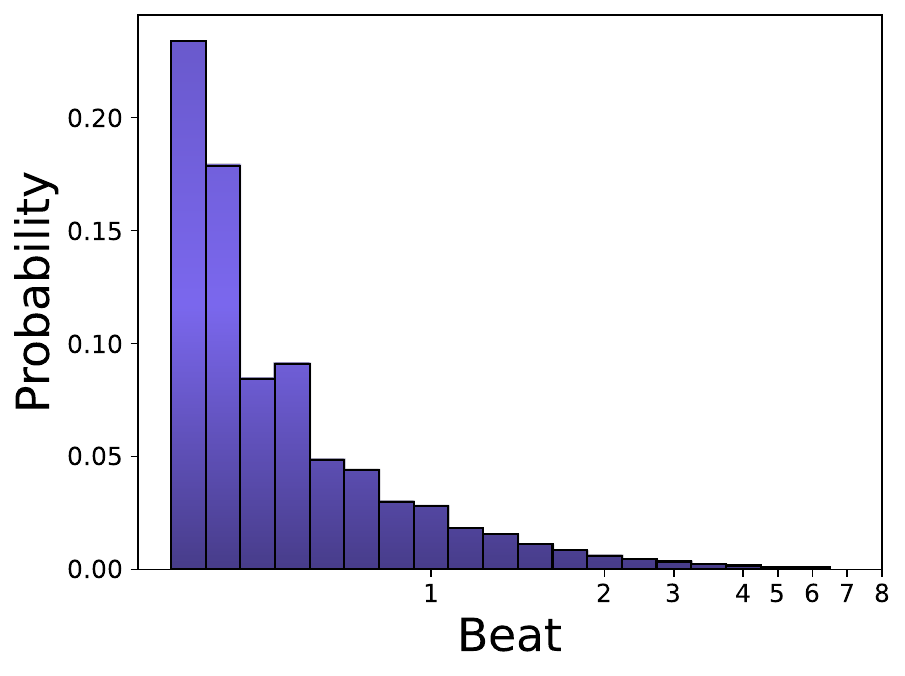}
    \vspace{-1.5\baselineskip}
    \caption[]{\textit{Ts} + \textit{NOff}}
  \end{subfigure}
  \begin{subfigure}{.195\textwidth}
    \centering
    \includegraphics[width=\textwidth]{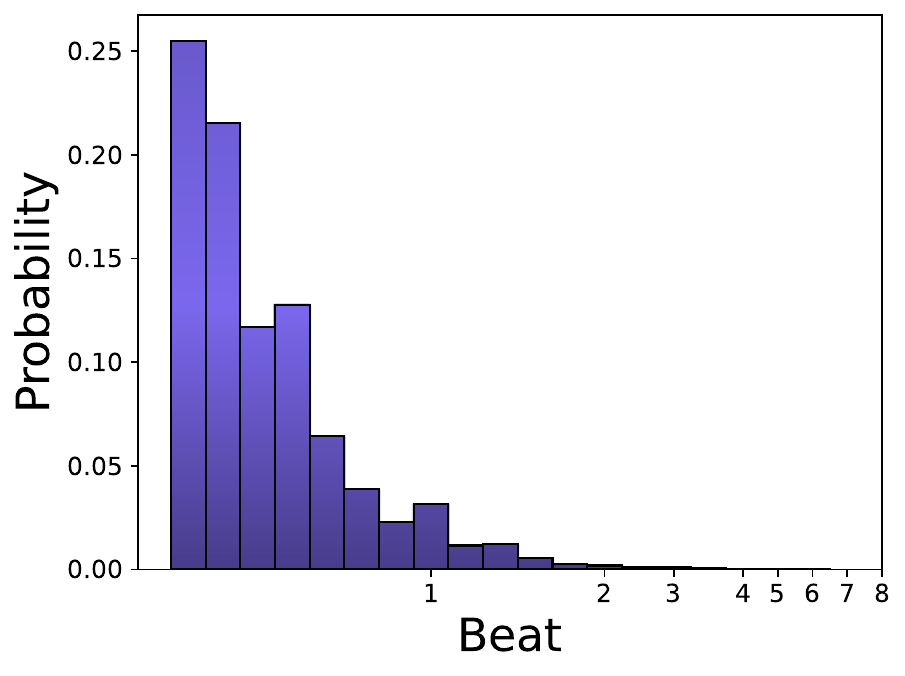}
    \vspace{-1.5\baselineskip}
    \caption[]{\textit{Pos} + \textit{Dur}}
  \end{subfigure}
  \begin{subfigure}{.195\textwidth}
    \centering
    \includegraphics[width=\textwidth]{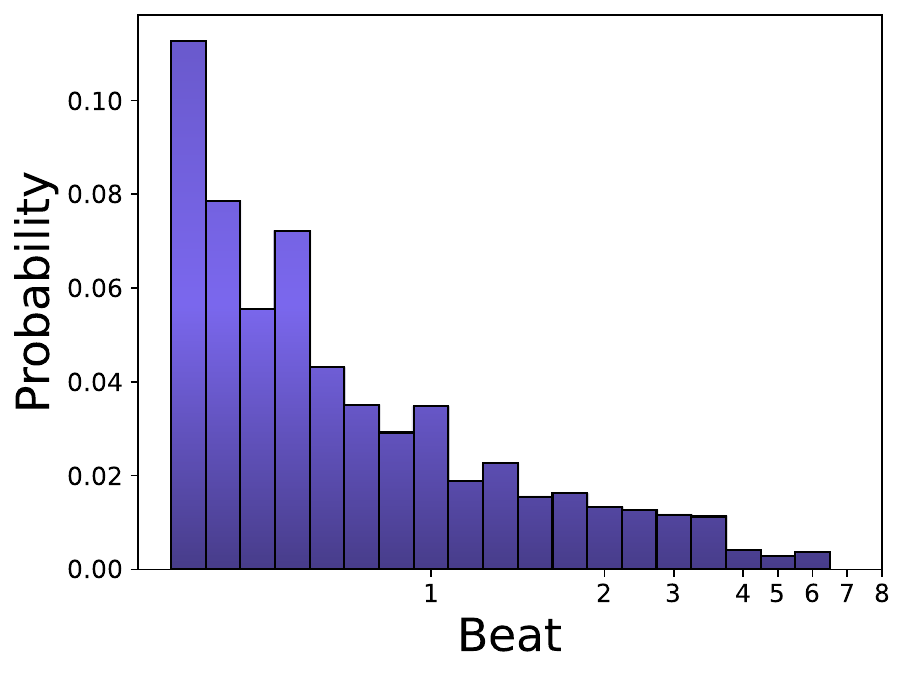}
    \vspace{-1.5\baselineskip}
    \caption[]{\textit{Pos} + \textit{NOff}}
  \end{subfigure}
  \begin{subfigure}{.195\textwidth}
    \centering
    \includegraphics[width=\textwidth]{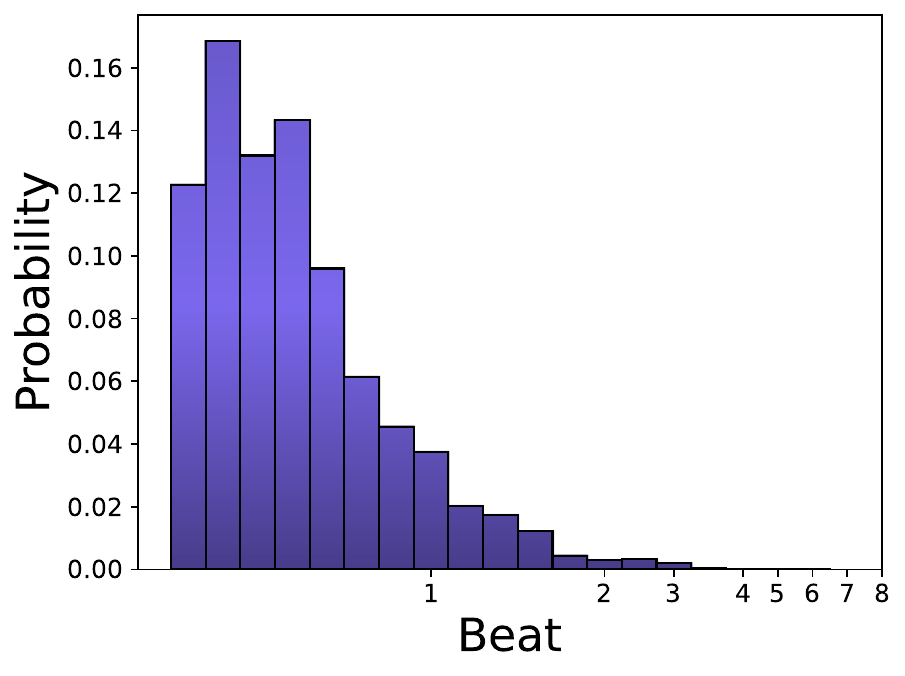}
    \vspace{-1.5\baselineskip}
    \caption[]{POP909 dataset}
  \end{subfigure}

  \caption{Histograms of the note onset positions within bars (top-row), note offset positions within bars (middle-row) and note durations (bottom-row) of the generated notes. There are 32 possible positions within a bar, numerated from 0 (beginning of bar) to 31 (last 32$^{th}$ note). The durations are expressed in beats, ranging from a 32$^{th}$ note to 8 beats.}
  \label{fig:tok_generation_positions_durations}
\end{figure*}

\begin{figure}[t]
\centering

  \begin{subfigure}{.495\columnwidth}
    \centering
    \includegraphics[width=\textwidth]{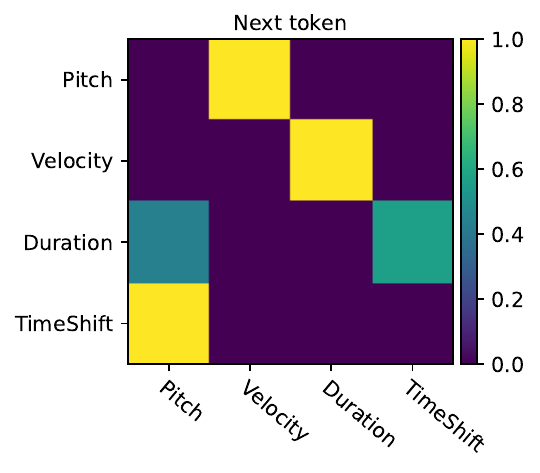}
    \vspace{-1.5\baselineskip}
    \caption[]{\textit{Ts} + \textit{Dur}}
  \end{subfigure}
  \begin{subfigure}{.495\columnwidth}
    \centering
    \includegraphics[width=\textwidth]{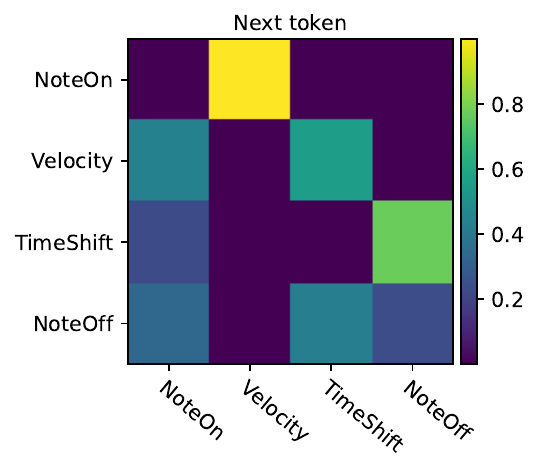}
    \vspace{-1.5\baselineskip}
    \caption[]{\textit{Ts} + \textit{NOff}}
  \end{subfigure}
  \begin{subfigure}{.495\columnwidth}
    \centering
    \includegraphics[width=\textwidth]{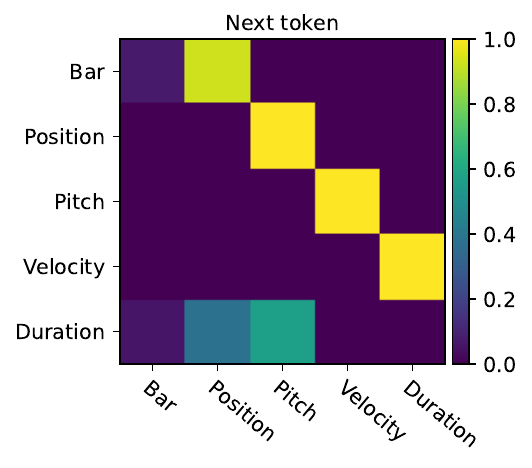}
    \vspace{-1.5\baselineskip}
    \caption[]{\textit{Pos} + \textit{Dur}}
  \end{subfigure}
  \begin{subfigure}{.495\columnwidth}
    \centering
    \includegraphics[width=\textwidth]{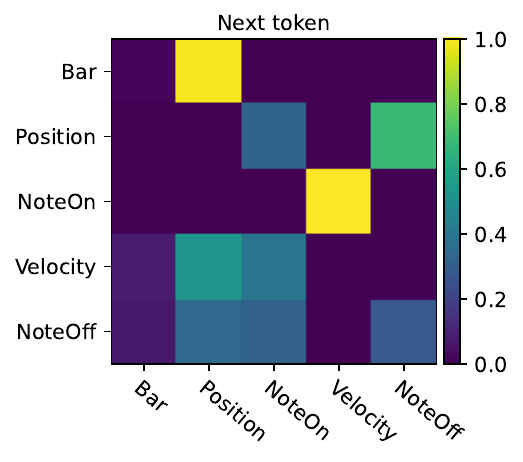}
    \vspace{-1.5\baselineskip}
    \caption[]{\textit{Pos} + \textit{NOff}}
  \end{subfigure}

  \caption{Token type succession heatmaps of the generated results. The vertical axis is a the current token type, the horizontal axis is the next token type following the current one. All rows are normalized.}
  \label{fig:tok_generation_token_successions}
\end{figure}

For the generative task, we use the POP909 dataset \cite{pop909-ismir2020}. The models start with prompt made of between 384 to 512 tokens, then autoregressively generate 512 additional tokens.
Evaluation of generated results remains an open issue \cite{on_the_eval_music}. Previous work often perform measures of similarity of certain features such as pitch range or class, between prompts and generated results, alongside human evaluations. Feature similarity is however arguably not very insightful: a generated result could have very similar features to its prompts while being of poor quality. Human evaluations, while being more reliable on the quality can also induce biases. Besides, \cite{huang_remi_2020} already shows results on an experiment similar to ours.

Hence we choose to evaluate results on the ratios of prediction errors: Token Syntax Error (TSE) \cite{bpe-for-symbolique-music2023}. This metric is bias-free and directly linked to the design choices of the tokenizations. It allows us to measure how a model achieves to make reliable predictions based on the input context and the knowledge it learned.

We use the categories from \cite{bpe-for-symbolique-music2023}:
\begin{itemize}
    \item $\mathbf{TSE_{type}}$: an error of type, e.g., when the model predicts a token of an incompatible type with the previous one.
    \item $\mathbf{TSE_{time}}$: a wrong predicted \texttt{Position} value, that goes back or stay in time.
    \item $\mathbf{TSE_{dupn}}$ (duplicated note): a note predicted whereas it was already being played at the current time being.
    \item $\mathbf{TSE_{nnof}}$ (no NoteOff): a \texttt{NoteOn} token been predicted with no following \texttt{NoteOff} token to end it.
    \item $\mathbf{TSE_{nnon}}$ (no NoteOn): \texttt{NoteOff} token predicted whereas this note was not being played.
\end{itemize}

For each generated token, a rule-based function analyzes its type and value to determine if both are valid, or which type of error was made otherwise. The overall number of errors is normalized by the number of predicted tokens.

\begin{table}
  \resizebox{\columnwidth}{!}{
  \begin{tabular}{lccccc}
    \toprule
    Tokenization & $\mathbf{TSE_{type}}$ $\downarrow$ & $\mathbf{TSE_{time}}$ $\downarrow$ & $\mathbf{TSE_{dupn}}$ $\downarrow$ & $\mathbf{TSE_{nnon}}$ $\downarrow$ & $\mathbf{TSE_{nnof}}$ $\downarrow$ \\
    \midrule
    \textit{TS} + \textit{Dur} & $<10^{-3}$ & - & 0.014 & - & - \\
    \textit{TS} + \textit{NOff} & $<10^{-3}$ & - & 0.001 & 0.109 & 0.040 \\
    \textit{Pos} + \textit{Dur} & 0.002 & 0.113 & 0.032 & - & - \\
    \textit{Pos} + \textit{NOff} & 0.002 & 0.127 & 0.005 & 0.095 & 0.066 \\
    \bottomrule
  \end{tabular}}
  \caption{Prediction error ratios when performing autoregressive generation. - symbol stands for not concerned, and can be interpreted as 0.}
  \label{tab:tok_gen_tse}
\end{table}

The results are reported in \cref{tab:tok_gen_tse}. We first observe that the type error ratios are lower than in other categories. This is excepted since it is less computationally demanding to model the possible next types depending solely on the last one, rather than on the value of the predicted token, for which the validity depends on a the whole previous context.

\texttt{Position} tokens bring almost no type errors, but a noticeable proportion of time errors. When decoding tokens to notes, this means that the time may go backward, and resulting in sections of overlapping notes.

Although \texttt{Duration} tokens seem to bring slightly more note duplication errors, the use of \texttt{NoteOn} and \texttt{NoteOff} tokens results in a considerable proportion of note prediction errors. \texttt{NoteOff} tokens predicted while the associated note was not being played ($\mathbf{TSE_{nnon}}$) does not have undesirable consequences when decoding tokens to notes, but it pointlessly extends the sequence, reducing the efficiency of the model, and may mislead the next token predictions. Additionaly, \texttt{NoteOn} tokens predicted without associated \texttt{NoteOff} ($\mathbf{TSE_{nnof}}$) result in notes not properly ended. This error can only be handled by applying a maximum note duration after decoding.
Explicit \texttt{Duration} tokens allows to specify in advance this information, for both short and long notes. Conversely, with \texttt{NoteOff} tokens, the note duration information is implicit and inferred by the combinations of \texttt{NoteOn}, \texttt{NoteOff} and time tokens. This can be interpreted as an extra effort for the model. Consequently, some uncertainty on the duration accumulates over autoregressive steps during generation.
Based on these results, the best tradeoff ensuring good predictions seems to represent time with \texttt{TimeShift} tokens and note duration with \texttt{Duration} tokens.


In \cref{fig:tok_generation_positions_durations} we observe the positions within bars and durations of the generated notes. In all cases, onset positions are more distributed at the beginning of the bars. This is especially the case with \texttt{Bar} and \texttt{Position} tokens, for which we may find unexpected rests at the end of bars, when \texttt{Bar} tokens are predicted during the generation before that the current bar is completed. The \textit{TS} + \textit{Dur} combination places note onsets much more on even positions. The probability mass of \textit{TimeShift} tokens (especially for short values) seems to be much higher. However, this is not the case for the \textit{TS} + \textit{NOff} combination, as \texttt{TimeShift} tokens have to be predicted to move the time on odd positions of note offsets. As shown in \cref{fig:tok_generation_token_successions}, right after the model is likely to predict a next note, resulting in evenly distributed onset distribution.

Finally, the use of \texttt{NoteOff} tokens tends to produce longer note durations, especially when combined with \texttt{Position} tokens. In this last case, we can assume that the model might "forget" the notes currently being played, and that it struggles more to model their durations that have to be implicitly deduced from the past \texttt{Bar} and \texttt{Position} tokens.


\begin{table}[h]
  \resizebox{\columnwidth}{!}{
  \begin{tabular}{lccc}
    \toprule
    Tokenization & Top-20 composers $\uparrow$ & Top-100 composers $\uparrow$ & Emotion $\uparrow$ \\
    \midrule
    \textit{TS} + \textit{Dur} & \textbf{0.973} & \textbf{0.941} & \textbf{0.983} \\
    \textit{TS} + \textit{NOff} & 0.962 & 0.930 & 0.962 \\
    \textit{Pos} + \textit{Dur} & 0.969 & 0.927 & 0.963 \\
    \textit{Pos} + \textit{NOff} & 0.963 & 0.925 & 0.956 \\
    \bottomrule
  \end{tabular}}
  \caption{Accuracy on classification tasks.}
  \label{tab:tok_classification}
\end{table}

\section{Classification}\label{sec:tok_classification}

For some classification tasks, symbolic music is arguably better suited than audio or piano roll. This is particularly true for classical music feature classification, such as composer \cite{kong2020largescale}. Mono-instrument music with complex melodies and harmonies and no particular audio effect benefit from being represented as discrete for classification and modeling tasks. Given this, it felt important to us to conduct experiments on such task.

We choose to experiment with the GiantMIDI \cite{giantMIDI2020} dataset for composer classification and the EMOPIA \cite{emopia} dataset for emotion classification.
The results, as shown in \cref{tab:tok_classification}, indicate that there is very little difference between the various tokenization methods. However, the combination of \texttt{TimeShift} and \texttt{Duration} consistently outperforms the others by one point

The classification task involves modeling the patterns from data that are characteristic to composers or emotions. Here, it seems that the time distance between notes, and their explicit duration play a role in these task, more than note offsets or onset positions. This comes with no surprise for the composer classification task, considering that the data is largely composed of complex music with dense melodies and harmonies, featuring mostly short successive notes. Intuitively, patterns of note successions and chords are more easily distinguishable with explicit durations. With implicit note durations, the overall patterns must be deduced by the combinations of \texttt{NoteOn} and \texttt{NoteOff} tokens while keeping track of the time.

\section{Sequence representation}\label{sec:tok_contrastive_learning}


\begin{figure*}
  \centering

  \begin{subfigure}{.48\textwidth}
    \centering
    \includegraphics[width=\textwidth]{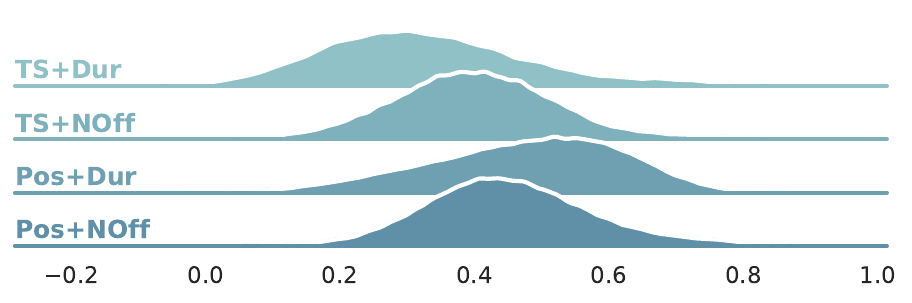}
    \vspace{-1.5\baselineskip}
    \caption[]{Pitch +1}
  \end{subfigure}
  \begin{subfigure}{.48\textwidth}
    \centering
    \includegraphics[width=\textwidth]{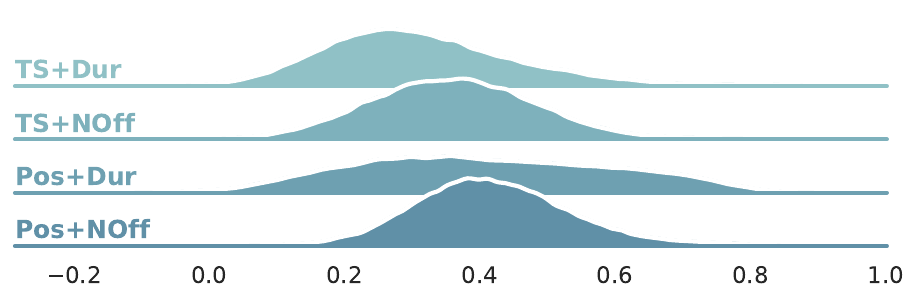}
    \vspace{-1.5\baselineskip}
    \caption[]{Pitch +12}
  \end{subfigure}

  \begin{subfigure}{.48\textwidth}
    \centering
    \includegraphics[width=\textwidth]{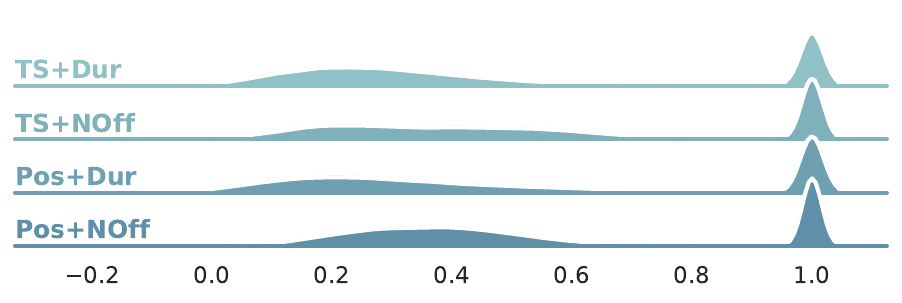}
    \vspace{-1.5\baselineskip}
    \caption[]{Velocity +1}
  \end{subfigure}
  \begin{subfigure}{.48\textwidth}
    \centering
    \includegraphics[width=\textwidth]{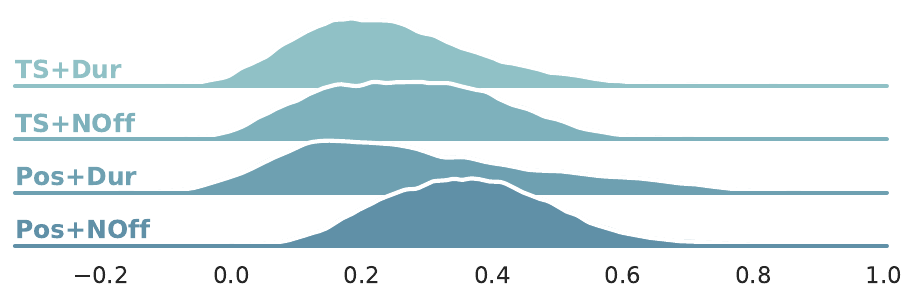}
    \vspace{-1.5\baselineskip}
    \caption[]{Pitch +12 and Velocity +1}
  \end{subfigure}

  \caption{Density plots of cosine similarities between pairs of original and augmented token sequences.}
  \label{fig:tok_contrastive_cosim}
\end{figure*}

The last task that we wished to explore is sequence representation. It consists in obtaining a fixed size embedding representation of an input sequence of tokens $p_\theta: \mathbb{V}^L \mapsto \mathbb{R}^d$. Here $\mathbb{V} \subset \mathbb{N}$ denotes the token ids of the vocabulary $\mathcal{V}$, $L$ is the variable input sequence length, and $d$ the size of embeddings. In other words, the model learns to project an input token sequence into a embedding space, thus providing a universal representation. We find this task interesting and well-suited to assess model performances as it directly trains it to model the relationships between tokens within the input sequence and between different representations themselves. While the real-world applications of this task for symbolic music are currently limited, it serves as a useful benchmarking technique for measuring how tokenization impacts the learning of models.

This task has previously been addressed in natural language processing by SentenceBERT \cite{reimers-2019-sentence-bert} or SimCSE \cite{gao2021simcse}. We adopted the approach of the latter, which uses contrastive learning to train the model to learn sequence representations, for which similar inputs have higher cosine similarities. The sequence embedding is obtained by performing a pooling operation on the output hidden states of the model. We decided to use the last hidden state of the \texttt{BOS} token position, as it yielded good results with SimCSE\cite{gao2021simcse}\footnote{SimCSE uses a \texttt{CLS} token which is equivalent to \texttt{BOS} in our case.}. We trained the models with the dropout method: during training, a batch of $n$ sequences $\mathcal{X} = \{ \mathbf{x}_i \}_{i=0}^n$ is passed twice to the model, but with different dropout masks, resulting in different output sequence embeddings $\mathcal{Z} = \{ \mathbf{z}_i \}_{i=0}^N$ and $\bar{\mathcal{Z}} = \{ \bar{\mathbf{z}_i} \}_{i=0}^N$. Although the dropout altered the outputs, most of the input information is still accessible to the model. Hence, we expect pairs of sequence embeddings $\left( \mathbf{z}_i, \bar{\mathbf{z}_i} \right)$ to be similar, so having a high cosine similarity. To achieve this objective, we train the model with a loss function defined by the cross-entropy for in-batch pairwise cosine similarities ($\mathrm{sim}$):

\begin{equation}\label{eq:tok_contrastive_objective}
  \ell_i = - \log \frac{e^{\mathrm{sim}(\mathbf{z}_i, \bar{\mathbf{z}_i}) / \tau }}{\sum_{j=1}^N e^{\mathrm{sim}(\mathbf{z}_i, \bar{\mathbf{z}_j}) / \tau }}
\end{equation}

As a result, the model will effectively learn to create similar sequence embeddings for similar inputs, while pushing apart those with dissimilarities. We kept a 0.1 dropout value to train the models, and used the GiantMIDI dataset \cite{giantMIDI2020}.

Evaluation of sequence representation is intuitively performed by measuring the distances and similarities of pairs of similar sequences. We resort to data augmentation by shifting the pitch and velocity of the sequences in order to get pairs of similar music sequences.
The augmented data keeps most of the information of the original data. As such, the models are expected to produce similar embeddings for pairs of original-augmented sequence. Ideally, the cosine similarity should be high, yet not to be equal to 1, as this would indicate that the model fails to capture the differences between the two sequences.
The results, presented in \cref{fig:tok_contrastive_cosim}, indicate that \texttt{Position}-based tokenizations perform slightly better. Therefore, it appears that explicit note onset and offset positions information facilitates models to obtains a universal musical representation.

Unlike classification, the contrastive learning objective models the similarities and dissimilarities between examples in the same batch. In this context, note onset and offset positions appear to be helpful for the models to distinguish music.

We also note the contrasting results when augmenting the velocity. Increasing it by one unit, which would be equivalent to playing just a little bit louder, have arguably a very small impact. As a result, the models mostly produces embeddings that are almost identical for the original and the augmented sequences, but also exhibits uncertainty for a notable proportion of samples.

To complement these results, we estimated the isotropy of sets of sequence embeddings. Isotropy measures the uniformity of the variance of a set points in a space. More intuitively, in an isotropic space, the embeddings are evenly distributed. It has been associated with improved performances in natural language tasks \cite{pmlr-v119-wang20k, bis-etal-2021-much, liang2021towards}, because embeddings are more equally distant proportionally to the density of their area, and are in turn more distinct and distinguishable. We choose to estimate it with the intrinsic dimension of the sets of embeddings. Intrinsic dimension is the number of dimensions required to represent a set of points. It can be estimated by several manners \cite{scikit-dimension}. We choose Principal Component Analysis (PCA) \cite{PCA}, method of moments (MOM) \cite{ID_MOM}, Two Nearest Neighbors (TwoNN) \cite{ID_TwoNN} and FisherS \cite{ID_fishers}. The results, reported in \cref{tab:tok_contrastive_id}, show that the embeddings created from the \textit{Pos} + \textit{Dur} combination tend to occupy more space across the dimension of the model, and are potentially better distributed.

\begin{table}[t]
  \resizebox{\columnwidth}{!}{
  \begin{tabular}{lcccc}
    \toprule
    Tokenization & lPCA $\uparrow$ & MOM $\uparrow$ & TwoNN $\uparrow$ & FisherS $\uparrow$ \\
    \midrule
    \textit{TS} + \textit{Dur} & \textbf{213} & 42.6 & 34.3 & 17.5 \\
    \textit{TS} + \textit{NOff} & 161 & 43.7 & 32.7 & 17.5 \\
    \textit{Pos} + \textit{Dur} & 146 & 39.1 & 33.1 & 17.1 \\
    \textit{Pos} + \textit{NOff} & 177 & \textbf{45.2} & \textbf{35.6} & \textbf{17.8} \\
    \bottomrule
  \end{tabular}}
  \caption{Intrinsic dimension of sequence embeddings, as an estimation of isotropy.}
  \label{tab:tok_contrastive_id}
\end{table}

\section{Conclusion}\label{sec:tok_conclusion}

We have discussed the importance of different aspects of symbolic music tokenization, and focused on two major ones: the time and note duration representations. We showed that different tokenization strategies can lead to different model performances due to the explicit information carried by tokens, depending on the task at hand.

Explicitly representing note duration leads to better classification accuracy as it helps the models to capture the melodies and harmonies of a music. Modeling durations, when represented implicitly, adds an extra effort to the model. However, the note offset position information it brings have been found to be more discriminative and effective in our contrastive learning experiment.

For music generation, the time representation plays a significant role, for which the note onset and offsets distributions vary due to the successions of token types. Implicit note durations are less suited for the autoregressive nature of this task, from a prediction error perspective, and sometimes "forgetting" notes being played resulting in higher durations.

We consider this work as a first step into the study of music tokenization for music modeling. We did not experiment with the other music tokenization dimensions, and other important tasks such as music transcription or synthesis for which we could find different results. For transcription, input audio frames are likely to contain the ending or beginning of notes, without being able to identify their onset or offset notes. Explicit \texttt{Duration} tokens might confuse models.
Furthermore, we believe that more musical reasoning tasks, imposing the model to perform logic deductions to retrieve implicit information from the data, might give more insightful results.
Future research will further explore these questions.

\bibliography{references}

\end{document}